\documentclass[12pt]{article}
\usepackage{graphicx}
\begin{document}
\title{\Large Inclusion of non-spherical components\\
 of the Pauli blocking operator\\
in (p,p$'$) reactions}
\author{E. J. Stephenson\\
\it{Indiana University Cyclotron Facility}\\
\it{Bloomington, IN 47408}\\ \\
R. C. Johnson\\
\it{Department of Physics, University of Surrey}\\
\it{Guildford, Surrey GU2 5XH, UK}\\ \\
F. Sammarruca\\
\it{Department of Physics, University of Idaho}\\
\it{Moscow, Idaho 83844}}
\date{\today}
\maketitle
\begin{abstract}
We present the first calculations of proton elastic and inelastic
scattering in which the Pauli blocking operator contains the
leading non-spherical components as well as the usual spherical
(angle-averaged) part. We develop a formalism for including the
contributions to the effective nucleon-nucleon interaction from
the resulting new $G$-matrix elements that extend the usual
two-nucleon spin structure and may not conserve angular momentum.
We explore the consequences of parity conservation, time reversal
invariance, and nucleon-nucleon antisymmetrization for the new
effective interaction. Changes to the calculated cross section and
spin observables are small in the energy range from 100 to
200~MeV.
\end{abstract}
% body of paper here
\section{Introduction}

Proton elastic and inelastic scattering at energies above about
100~MeV are usually described by distorted-wave calculations based
on an effective nucleon-nucleon (NN) interaction.  When the
interaction with the projectile proton is summed, or ``folded,''
over all the nucleons in the target, the resulting potential can
be used to model elastic scattering. In addition, the same
effective NN interaction becomes the transition potential in the
Distorted Wave Impulse Approximation (DWIA) that connects to
excited states of the target when the struck nucleon moves to a
new shell-model orbit, creating a particle-hole pair.  The
many-body effects of the nuclear medium are usually incorporated
through modifications to this effective NN interaction that depend
on the local nuclear density.

Systematic studies \cite{r1,r2,r3,r4,r5} have shown that one
important contribution to the many-body effects is Pauli blocking,
particularly at the lower end of the intermediate energy range.
This is the mechanism which prevents nucleons in the nuclear
medium from scattering to occupied intermediate states \cite{r6}.
This restriction is included through a projection operator in the
Bethe-Goldstone equation for the $G$-matrix elements that describe
the effective NN interaction inside the nuclear medium.

The usual practice is to average the Pauli projection operator
over the intermediate state scattering angle. If, instead of this
``spherical'' approximation, the non-spherical components are
retained, new $G$-matrix elements appear \cite{r7,r8,r9,r10}.
While remaining diagonal in total spin $S$ and isospin $T$, the
angular dependence allows coupled $G$-matrix elements that connect
partial wave states where $J\neq J'$ and $\ell\neq\ell '$ beyond
the $\vert\ell -\ell '\vert =2$ coupling generated by the tensor
interaction. The expanded set of $G$-matrix elements also depends
on $M$, the magnetic quantum number of the total angular momentum
$J$.  In studies that considered the effects on nuclear binding
energies, small but non-negligible changes were found when these
modifications to the $G$-matrix were included \cite{r8,r9}.  It
has also been suggested that the spherical approximation is
adequate for the central and spin-orbit parts of the effective
interaction \cite{r7}, those pieces that matter the most for
elastic scattering and the excitation of natural-parity
transitions.  But no calculations have been made to check
quantitatively in a comparison with scattering measurements how
important it is to treat the non-spherical Pauli blocking
components.

In a recent calculation of the non-spherical $G$-matrix
\cite{r10}, we observed that the new couplings in $J$ and $\ell$
generated only small matrix elements, but the variation of the
$G$-matrix elements with respect to $M$, the projection of the
total angular momentum $J$, was comparable to the typical size of
conventional medium effects. Thus it seemed appropriate to
investigate whether these modifications could have an impact on
nuclear reactions, and in particular on their spin observables,
which are most sensitive to the non-spherical components of the
nuclear force. Proton scattering offers a rich set of polarization
observables, including polarization transfer, that reflect the
spin dependence of the effective nucleon-nucleon interaction
itself.

The dependence on $M$ has been considered previously in
calculations of deuteron binding when the deuteron is treated as a
projectile traveling through the nuclear medium \cite{r11}. While
Pauli blocking reduces the deuteron binding energy as the density
increases, the amount depends on whether the projection of the
deuteron's spin, $\vert M\vert$, is 0 or 1 when the quantization
axis is taken to be along the direction of motion of the deuteron
through nuclear matter.  This difference generates a $T_P$-type
momentum-dependent tensor potential in the deuteron-nucleus
optical model.  Searches for such a potential have not been
definitive because of interpretive complications arising from the
strong coupling to breakup states \cite{r13}.

In this paper, we begin with the $G$-matrix described in
Ref.~\cite{r10}. The important features of the calculation of the
$G$-matrix elements are reviewed in Section~II.  In order to use
this effective interaction in distorted-wave calculations with
presently available computer programs, we must transform the
$G$-matrix elements to a coordinate-space representation using a
sum of Yukawa functions.  In Ref.~\cite{r5} these coefficients in
the Yukawa expansion are fit directly to the values of the
$G$-matrix elements.  This is equivalent to earlier methods in
which the nucleon-nucleon scattering amplitudes were calculated as
a function of scattering angle or momentum transfer and then the
Yukawa expansion ranges and coefficients were chosen to best
reproduce these angular distributions \cite{r14}.  For technical
reasons, we will follow the second scheme here (see Section III).
In the process, we will introduce a multipole expansion of the new
$G$-matrix elements in which the lowest order recovers the result
for a spherical Pauli blocking operator. For the spherical Pauli
operator, a comparison of the two methods shows practically
identical answers.

In Section IV we will compare the non-spherical and spherical
treatments of the Pauli blocking operator for representative
nuclear transitions at 100 and 200~MeV.  We will include a
comparison to the free, or density-independent, effective
interaction and to an effective interaction that also contains
relativistic effects. These comparisons will help gauge the
importance of the non-spherical treatment of Pauli blocking
relative to density-dependent effects more typically included. We
will also include measurements of the cross section and analyzing
power to illustrate these effects in relation to the quality of
the reproduction of the data. Pauli blocking has its largest
effects on the isoscalar central and spin-orbit terms in the
effective interaction.  These terms are well tested by a
comparison to elastic proton scattering or transitions to
natural-parity excited states.

Pauli blocking is only one process that is important in the
calculation of the effective interaction in the nuclear medium.
Others, such as the effects of strong relativistic mean fields
\cite{r10,r15,r16} and coupling to $\Delta$-resonances \cite{r17},
increase the repulsion in the nuclear medium just as does Pauli
blocking \cite{r18}. In a complete treatment, these should be
properly considered, along with the attraction expected to arise
from many-body forces. Thus a set of calculations based on the
conventional Brueckner-Hartree-Fock approach to nuclear matter
only (such as those we present here), should not be expected to
provide the final answer. However, a critical evaluation of any of
these medium effects requires that the treatment of Pauli blocking
not introduce systematic errors large enough to affect our
interpretation when agreement with data is considered. We will
show that at intermediate energies the inclusion of non-spherical
components in the blocking operator produces changes that are
modest in size compared to the main density-dependent effects
typically included in the effective interaction.

\section{Calculation of the $G$-matrix elements}

The Brueckner-Bethe-Goldstone equation \cite{r19,r20,r21,r22}
describes the scattering of two nucleons in nuclear matter. The
presence of the (infinite) nuclear medium is included through
Pauli blocking and a mean field arising from the interactions with
all the other nucleons. It is convenient to express the momenta of
the two nucleons, ${\bf k}_1$ for the projectile and ${\bf k}_2$
for the struck nucleon, in terms of the relative and
center-of-mass motion as \hfill\break
\parbox{14.6cm}{\begin{eqnarray*}
{\bf k}&=&({\bf k}_1-{\bf k}_2)/2\\
{\bf P}&=&({\bf k}_1+{\bf k}_2)~.
\end{eqnarray*}}\hfill\parbox{1cm}{\begin{eqnarray}\end{eqnarray}}
The total or center-of-mass
momentum {\bf P} is conserved in the scattering process.

In analogy with free-space scattering, the nuclear matter
Bethe-Goldstone equation is given by
\begin{equation}G({\bf k}',{\bf k},{\bf P},E_0) = V({\bf k}',{\bf k})\\
+ \int {d^3{\bf k}''\over(2\pi)^3}V({\bf k}',{\bf k}'') {Q({\bf
k}'',{\bf P})\over E_0+\imath \epsilon -E({\bf P},{\bf k}'')}
G({\bf k}'',{\bf k},{\bf P},E_0)\ ,\end{equation} where $V$ is the
two-body potential. The energy of the two-particle system, $E$
(with $E_0$ its initial value), includes kinetic energy and the
potential energy generated by the mean field. The latter is
determined in a separate self-consistent calculation of nuclear
matter properties and is conveniently parametrized in terms of
effective masses [5]. At this point the direction and magnitude of
{\bf P} have not been specified.

The Pauli projection operator $Q$ selects intermediate states
where both nucleon momenta lie above the Fermi momentum $k_F$:
\begin{equation}Q({\bf k},{\bf P},k_F)=\cases{1&if $k_1,k_2>k_F$\cr
0&otherwise\cr}\end{equation} Visualizing the (sharp) Fermi
surface as a sphere of radius $k_F$, the condition above imposes
the requirement that the tips of the ${\bf k}_1$ and ${\bf k}_2$
vectors lie outside the sphere. For applications to real nuclei,
$k_F$ is treated as a function of the local nuclear density
\begin{equation}\rho(r)={2k_F^3(r)\over 3\pi^2}\ .\end{equation}

The matrix elements for the non-spherical Pauli operator may be
written in a partial wave basis as\hfill\break
\parbox{14.6cm}{\begin{eqnarray*}\lefteqn{\langle
(\ell'S)J'M'\vert Q({\bf k},{\bf P},k_F)\vert (\ell S)JM\rangle}\\
&=&\sum_{m_\ell ,m_{\ell '},m_S} \langle\ell 'm_{\ell '}Sm_S\vert
J'M'\rangle\langle JM\vert\ell m_\ell Sm_S\rangle
\langle\ell'm_{\ell '}\vert Q({\bf k},{\bf P},k_F)\vert\ell
m_\ell\rangle\ ,\end{eqnarray*}}\hfill\parbox{1cm}{\begin{equation}
\end{equation}}
where \begin{equation}\langle
\ell'm_{\ell '}\vert Q({\bf k},{\bf P},k_F)\vert\ell m_\ell\rangle
=\int d\Omega\ Y^*_{\ell'm_{\ell '}} (\Omega )\ Y^{}_{\ell
m_\ell}(\Omega )\ \Theta (\vert{\bf k}_1\vert -k_F)\Theta (\vert
{\bf k}_2\vert -k_F)\ .\end{equation}  The step functions destroy
the orthogonality that would otherwise exist for the spherical
harmonics in the integral.  This allows couplings where
$\ell\neq\ell '$ and, through the recoupling coefficients in the
summation of Eq.~(5), couplings where $J\neq J'$.

If the quantization axis for the projection quantum numbers in
Eq.~(6) is chosen to lie along ${\bf P}$, then there is no
dependence on azimuthal direction in the $\Theta$ functions, and
integration over $d\Omega$ gives $m_\ell =m_{\ell '}$. Because
$Q({\bf k},{\bf P}, k_F)$ is diagonal in $m_S$, Eq.~(5) gives
$M=M'$. It then follows from Eq.~(2) that $G$ is diagonal in $M$.

Eq.~(5) can be further reduced using standard angular momentum
algebra to give \hfill\break
\parbox{14.6cm}{\begin{eqnarray*}\lefteqn{\langle
(\ell'S)J'M\vert Q({\bf k},{\bf P},k_F)\vert (\ell S)JM\rangle}\\
& = & {1\over 2}\sum_{L_1}(-1)^{\ell +S+J}\hat{J}'\hat{L}_1^2
\hat{\ell}' \ \langle\ell '0,L_10\vert\ell 0\rangle\ W(J'J\ell
'\ell ; L_1S)\\ & & \times\ 
\langle J'M,L_10\vert JM\rangle\ \theta
_{L_1}(P,k,k_F)\end{eqnarray*}}\hfill\parbox{1cm}{\begin{eqnarray}
\end{eqnarray}}
where $\theta _{L_1}(P,k,k_F)$ is the combination of step
functions defined in Eqs.~(26)-(27) of Ref.~\cite{r11}. Here and
elsewhere we use the notation $\hat X=\sqrt{2X+1}$. In Eq.~(7) the
quantum number $M$ is the projection of ${\bf J}$ along the
direction of ${\bf P}$. We have checked that values of selected
matrix elements calculated with this formula and Eq.~(6) agree
satisfactorily.

The $G$-matrix elements are obtained by solving the integral
Bethe-Goldstone equation, Eq.~(2), in the $\ell$, $S$, $J$, $M$,
$T$ basis. For any choice of {\bf P}, the $G$-matrix elements,
$\langle\ell 'J'M'\vert G^{ST}({\bf P})\vert\ell JM \rangle$, may
be expanded in spherical harmonics $Y_{L\Lambda}({\bf P})$ where
$L$ is the angular momentum that recouples $J$ to $J'$.  The
coefficients in the recoupling expansion are $G^{LTS}(\ell
'J',\ell J, P)$. Thus, \begin{equation}\langle\ell'J'M'\vert
G^{ST}({\bf P}) \vert\ell JM\rangle
=\sqrt{4\pi}\sum_{L\Lambda}\langle J'M',L\Lambda\vert JM\rangle\
\hat L \ Y_{L\Lambda}({\bf P})\ G^{LTS}(\ell 'J',\ell J,P)\
,\end{equation} where the arguments of spherical harmonics are
always unit vectors in the direction of the vector indicated. In
the limit of a spherically averaged Pauli blocking operator only
the $L=0$ terms in Eq.~(8) survive. Equation (8) can be inverted
to yield the expansion coefficients \begin{equation}G^{LTS}(\ell
'J',\ell J,P) ={1\over\hat J^2}\sum_M\langle J'M,L0\vert
JM\rangle\ \langle\ell 'J' M\kern-3pt :\kern-3pt{\bf P}\vert
G^{ST} \vert\ell JM\kern-3pt :\kern-3pt{\bf P}\rangle\
.\end{equation} where the relation $M=M'$ has been applied. The
$G$-matrix elements on the right hand side are defined with
respect to a basis $\vert\ell JM\kern-3pt :\kern-3pt{\bf
P}\rangle$
 and $M$ refers to the direction of {\bf P}.
For the nuclear matter calculations described here the magnitude
of {\bf P} has been chosen to be $k_1$.

Note that if the matrix elements $\langle\ell 'J' M\kern-3pt
:\kern-3pt{\bf P}\vert G^{ST} \vert\ell JM\kern-3pt :\kern-3pt{\bf
P}\rangle$ are independent of $M$ and diagonal in $J$ (spherically
averaged Pauli blocking) the right hand side of Eq.~(9)
automatically vanishes unless $L=0$ because of a property of the
C-G coefficients: \begin{displaymath}\sum_M\langle JM,L0\vert
JM\rangle=\hat J ^2 \delta _{L0}.\end{displaymath}

The anti-symmetrized $G$-matrix elements are obtained by
subtracting from Eq.~(9) the same terms but with an additional
phase of $(-)^{\ell + S+T}$.  This projects out the set of matrix
elements with $\ell +S+T$ even, leaving the set that normally
describes NN scattering. Due to parity conservation, $L$ takes on
only even values. The first step in our transformation to
coordinate space was to obtain these $L$-dependent expansion
coefficients.  This expansion is expected to converge quickly in
$L$, and the $L=0$ part was checked for consistency with the
result for the spherically averaged Pauli operator.

The $G$-matrix elements that we will use here were generated from
a free-space NN interaction that is a modified version of the
Bonn-B potential \cite{r18}. Details can be found in
Ref.~\cite{r5}. The model parameters were adjusted so as to
achieve a good reproduction of the phase shift analysis results
from the Nijmegen group \cite{r23} up to 325~MeV.

The number of matrix elements $\langle\ell 'J' M\kern-3pt
:\kern-3pt{\bf P}\vert G^{ST} \vert\ell JM\kern-3pt :\kern-3pt{\bf
P}\rangle$ that are coupled  in Eq.~(2) increases with $M$ as it
becomes possible to incorporate larger values of $J$.  However,
the $M$-dependence decreases with increasing $J$ \cite{r10} and we
were able to ignore these effects on partial waves with $J,J'>6$,
which were calculated using the usual angle-average approximation.

In general, the new elements of the $G$-matrix will introduce spin
operators that depend on the direction of ${\bf P}$, the sum of
the projectile momentum ${\bf k}_1$ and the momentum ${\bf k}_2$
of the particle encountered in the nuclear medium.  For practical
reasons it is desirable that ${\bf P}$ be fixed for a particular
transition and incident energy. At  the intermediate energies of
interest here, when the incident laboratory momentum can be
considered large compared with the momenta of the target nucleons,
a natural approximation for  ${\bf P}$ is ${\bf k}_1$. We shall
show however that this is not a good choice  here because it leads
to a G-matrix which cannot be expressed in terms of the standard
set of spin operators (for example, $\sigma_1\cdot\sigma_2$,
$(\sigma_1+\sigma_2)\cdot\hat n$, $S_{12}(\hat q)$, and
$S_{12}(\hat Q)$, where the momentum transfer $\hat q$ and the
normal to the scattering plane $\hat n$ establish a coordinate
system with $\hat Q=\hat q\times\hat n$). In this paper we choose
${\bf P}$ to be parallel to ${\bf Q}$ and have magnitude $k_1$.
This does lead to a G-matrix with the standard spin structure. Our
choice for the magnitude is a standard one in calculations of
medium effects.

\section{Transformation of the $G$-matrix to coordinate\\ space}

A calculation of proton elastic or inelastic scattering
observables using presently available computer programs requires
that the information on the NN interaction in the medium carried
by the $G$-matrix elements be converted into the amplitudes of the
effective NN interaction.  Usually, the amplitudes for each spin
and isospin operator are expanded in a Yukawa series as a function
of the momentum transfer.  We will begin this section with a
description of the general formulas that connect the expanded
$G$-matrix of Eqs.~(8) and (9) to NN scattering amplitudes,
regardless of the complexity of the coupling.  This makes them
available for any possible future application.  For the
calculations we show here, we will restrict ourselves to the set
of operators associated with free NN scattering since these are
the conventionally used set. After the general formulas, we will
show the forms actually used for our calculations.

We first convert from the $\ell, S, J,M,T$ basis to a
representation characterised by definite incoming and outgoing
momenta and intrinsic spin projections.  The amplitudes in the two
bases are related by\hfill\break\parbox{14.6cm}{\begin{eqnarray*}
\langle{\bf k}',S\sigma '\vert G^T_{}({\bf P})\vert{\bf
k},S\sigma\rangle &=&(2\pi )^3\ \sum_\zeta\ \imath^{\ell -\ell
'}\langle\ell 'm_{\ell '},S\sigma ' \vert J'M'\rangle\langle\ell
m_\ell , S\sigma\vert JM\rangle\\ &\times &\langle J'M',
L\Lambda\vert JM\rangle\ Y^{}_{\ell 'm_{\ell '}} ({\bf k'})\
Y^{*}_{\ell m_\ell}({\bf k})\\ &\times &
\hat L\sqrt{4\pi}\ Y_{L\Lambda}({\bf
P})\ G^{LTS}_{}(\ell 'J',\ell J,P)\ ,\end{eqnarray*}}\hfill
\parbox{1cm}{\begin{eqnarray}\end{eqnarray}}
with a summation that runs over $\zeta =\ell 'J'M'\ell JMm_{\ell
'}m_\ell L\Lambda$. The momenta ${\bf k}$ and ${\bf k'}$ are
nucleon momenta in the NN center-of-mass system [see Eq.~(1)].

With the normalization factors as in Eq.~(10), the partial wave
matrix elements $G^{LST}(\ell 'J,\ell J,P)$ satisfy unitarity
relations which, in the case of uncoupled partial waves in free
space, have the form \begin{equation}\Im\Biggl({1\over G}\Biggr)
={\mu k\over 8\pi^2\hbar^2}\end{equation} where $\mu$ is the NN
reduced mass. This relation is equivalent to
\begin{equation}G_\ell =-{8\pi^2\hbar^2\over\mu k}\ \exp
(i\delta_\ell\sin\delta_\ell )\end{equation} where $\delta_\ell$
is a real NN phase shift.  This is the same definition of the
partial wave $G$-matrix elements that was used in Appendix~A of
Ref.~\cite{r5}.

We denote antisymmetrized matrix elements by $\tilde G^T$ where
\begin{equation}\langle{\bf k}',S\sigma '\vert\tilde G^T({\bf P})\vert{\bf k},
S\sigma\rangle =2\ \langle{\bf k}',S\sigma '\vert G^T({\bf
P})\vert {\bf k},S\sigma\rangle\ .\end{equation} For this to be
correct, the summation of Eq.~(10) and subsequent summations in
this paper must restrict $\ell$ and $\ell '$ to satisfy
\begin{equation}(-1)^{\ell +S+T}=(-1)^{\ell' +S+T}=-1\
.\end{equation}

For the purpose of applications in DWBA calculations
 we wish to separate the parts of $\tilde G$
associated with the central, spin-orbit, and tensor operators
usually used to describe the spin structure of the on-shell
($k=k'$) NN scattering amplitude. In particular, we want to
consider the form \begin{displaymath}\langle{\bf k}',S\sigma
'\vert\tilde G^{T}_{}({\bf P})\vert{\bf k},S\sigma\rangle
=\sum_L\langle{\bf k}',S\sigma '\vert\tilde G^{LT}({\bf
P})\vert{\bf k},S\sigma\rangle,\end{displaymath} where\hfill\break
\parbox{14.6cm}{
\begin{eqnarray*}\lefteqn{\langle{\bf k}',S\sigma '\vert \tilde
G^{LT}_{}({\bf P})\vert{\bf k},S\sigma\rangle = \tilde
G^{LTS}_C(\theta )}\\ & & + 
\delta_{S1}\bigl[\tilde G^{L1T}_{LS}(\theta )
(\vec\sigma_1+\vec\sigma_2) \kern-2pt\cdot\kern-2pt\hat n+\tilde
G^{L1T}_{TD}(\theta )S^{}_{12}(\hat q)+\tilde G^{L1T}_{TX}(\theta
) S^{}_{12}(\hat Q)\bigr]\ ,\end{eqnarray*}} \hfill\parbox{1cm}
{\begin{equation}\end{equation}}
and where $\vec {\bf
q}=\vec{\bf k}'_1-\vec{\bf k}_1=\vec{\bf k}'-\vec{\bf k} $ is the
momentum transfer, $\hat n$ is the normal to the scattering plane,
and $\hat Q=\hat q\times\hat n$. Each of the $\tilde G^{LTS}_i$ is
a function of the scattering angle $\theta$. The subscript
indicates the spin operator in the NN amplitude corresponding to
that coefficient, using $C$ for central (both $S=0$ and $S=1$),
$LS$ for spin-orbit, and $TD$ and $TX$ for the ``direct'' and
``exchange'' parts of the tensor interaction.

Quite generally,  the $G$-matrix $\langle{\bf k}',S\sigma
'\vert\tilde G^{LT}_{}({\bf P})\vert{\bf k},S\sigma\rangle $ can
be expanded in terms of the complete set of spin tensors
$\tau_{k_Sq_S}(S)$  whose matrix elements are
\begin{equation}\langle S\sigma '\vert\tau_{k_Sq_S}\vert
S\sigma\rangle = \hat k_S\ \langle S\sigma ,k_Sq_S\vert S\sigma
'\rangle ,\end{equation} where $k_S$ runs from 0 to $2S$. The
coefficients in this expansion are $\tilde G^{LTS}_{k_Sq_S}({\bf
k},{\bf k}')$ and are given by \begin{equation}\tilde
G^{LTS}_{k_Sq_S}({\bf k},{\bf k}',{\bf P})={\rm Trace}\
\bigl(\langle{\bf k}',S\sigma '\vert\tilde G^{LT}_{}({\bf
P})\vert{\bf k},S\sigma\rangle\ \tau_{k_Sq_S}\bigr)\
.\end{equation} These coefficients $\tilde G^{LTS}_{k_Sq_S}({\bf
k},{\bf k}', {\bf P})$ are defined and normalized so that
\begin{equation}\langle{\bf k}',S\sigma '\vert\tilde G^{LT}_{}({\bf
P})\vert{\bf k},S\sigma\rangle=\hat{S}^{-2}\sum_{k_Sq_S} \langle
S\sigma '\vert\tau_{k_Sq_S}^{\dagger }\vert S\sigma\rangle\ \tilde
G^{LTS}_{k_Sq_S}({\bf k}, {\bf k}', {\bf P}). \end{equation}

We emphasize that for a general ${\bf P}$ the amplitude
$\langle{\bf k}',S\sigma '\vert\tilde G^{LT}_{}({\bf P})\vert{\bf
k},S\sigma\rangle$ of Eq.~(13) does not have the form of Eq.~(15)
but will contain other terms, {\it e.g.}, $S_{12}({\bf P})$.
However, it is shown in Appendix~A that if ${\bf P}$ is chosen so
that it becomes ${\bf P}'=\pm {\bf P}$ under a rotation of $\pi $
about $({\bf k}+ {\bf k'})$, then  the form Eq. (15) is
sufficiently general. The following formulas will assume that this
is the case. The coefficients in Eq.~(15) are then given in terms
of the $\tilde G^{LTS}_{k_Sq_S}$ by \begin{eqnarray}\tilde
G^{LTS}_C(\theta )&=&\hat S^{-2}\ \tilde G^{LTS}_{00} ({\bf
k},{\bf k}',{\bf P})\\ \tilde G^{L1T}_{LS} (\theta
)&=&{\sqrt{2\pi}\over 6}\sum_{q_S}Y^*_{1q_S}({\bf n})\ \tilde
G^{L1T}_{1q_S}({\bf k},{\bf k}',{\bf P})\\ \tilde G^{L1T}_{TD}
(\theta )-{1\over 2}\tilde G^{L1T}_{TX}(\theta )&=&{\sqrt{10\pi}
\over 30}\sum_{q_S}Y^*_{2q_S}({\bf q})\ \tilde G^{L1T}_{2q_S}({\bf
k},{\bf k}',{\bf P})\\ -{1\over 2}\tilde G^{L1T} _{TD}(\theta
)+\tilde G^{L1T}_{TX}(\theta ) &=&{\sqrt{10\pi}\over
30}\sum_{q_S}Y^*_{2q_S}({\bf Q})\ \tilde G^{L1T}_{2q_S}({\bf k},
{\bf k}',{\bf P})\ .\end{eqnarray} These results follow from the
orthogonality of the $\tau_{k_Sq_S}$ with respect to the trace
operation and the formulas ($S=1$) \begin{eqnarray}{\rm
Trace}(\tau_{1q_S}{\bf S}\kern-2pt\cdot\kern-2pt {\bf
A})&=&\sqrt{8\pi }\  Y_{1q_S}({\bf A})\\
 {\rm Trace}(\tau_{2q_S}(3({\bf S}\kern-2pt\cdot\kern-2pt
{\bf A})^2-2))&=&3\sqrt{8\pi \over 5}\ Y_{2q_S}({\bf A})\
,\end{eqnarray}
 where $\vec{A}$ is an arbitrary unit vector.

Using the partial wave expansion given in Eq.~(10) the trace  in
Eq.~(17) can be calculated explicitly to give,
\hfill\break\parbox{14.6cm}{\begin{eqnarray*}\tilde
G^{LTS}_{k_Sq_S}({\bf k}, {\bf k}',{\bf P})&=&\sum_\zeta
\imath^{\ell -\ell'}\hat S\hat J^2\hat J'\hat k_\ell\hat
L(-1)^{J-J'-\ell +k_\ell} \left\{ \matrix{\ell &\ell '&k_\ell\cr
J&J'&L\cr S&S&k_S\cr}\right\}\\ 
& \times &\langle k_\ell q_\ell
,L\Lambda\vert k_Sq_S\rangle\ \langle\ell 'm_{\ell
'},\ell m_\ell \vert k_\ell q_\ell\rangle\\ 
& \times & Y_{\ell 'm_{\ell
'}}({\bf k'})\ Y_{\ell m_\ell}({\bf k})\ \sqrt{4\pi}\
Y_{L\Lambda}({\bf P})\ \tilde G^{LTS}_{}(\ell 'J',\ell J,P)
\end{eqnarray*}}\hfill\parbox{1cm}{\begin{eqnarray}\end{eqnarray}}
where the sum runs over $\zeta =\ell 'J'\ell Jm_{\ell
'}m_\ell k_\ell q_\ell\Lambda$ with the restrictions for
antisymmetrization noted above.

For our present application, we  choose to express the components
of these tensors in a right-handed coordinate system with $\hat z$
along {\bf k} and $\hat y$ along $\hat n$. We call this the
`standard' coordinate system. We choose ${\bf P}$ to be parallel
to $({\bf k}+ {\bf k'})$ so that the restricted form of Eq.~(15)
is valid. In the limit that the reaction Q-value is small compared
to the bombarding energy, the direction of {\bf P} is toward
$\theta /2$. We also neglect terms with $J'\neq J$ because these
matrix elements turn out to be small. Eq.~(25) reduces to
\hfill\break \parbox{14.6cm}{\begin{eqnarray*}\tilde
G^{LTS}_{k_Sq_S}({\bf k}, {\bf k}', {\bf P}\vert \vert {\bf k}+
{\bf k}')&=&\sum_\zeta \imath^{\ell -\ell'}\hat S\hat J^3\hat
k_\ell\hat L(-1)^{-\ell +k_\ell} \left\{ \matrix{\ell &\ell
'&k_\ell\cr J&J&L\cr S&S&k_S\cr}\right\}\\ 
& \times &\langle k_\ell q_\ell
,L\Lambda\vert k_Sq_S\rangle\ \langle\ell '
q_\ell,\ell 0 \vert k_\ell q_\ell\rangle\\ 
& \times & Y_{\ell
'q_\ell}(\theta,0)\hat \ell \ Y_{L\Lambda}(\theta /2,0)\ \tilde
G^{LTS}_{}(\ell 'J,\ell J,P)\ ,\end{eqnarray*}}\hfill\parbox{1cm}
{\begin{eqnarray}\end{eqnarray}} The values of $\tilde
G^{LTS}(\ell 'J,\ell J,P)$ were calculated using Eq.~(9) in the
form\hfill\break \parbox{14.6cm}{\begin{eqnarray*}
\lefteqn{\tilde G^{LTS}(\ell J,\ell J,P)={1\over \hat
J^2}\ \Bigl\{\langle J0,L0\vert J0\rangle\ \langle\ell '
JM=0\vert\tilde G^{ST}\vert \ell JM=0\rangle}\\
& & +2\sum_{M>0}\langle
JM,L0\vert JM\rangle \ \langle\ell 'JM\vert\tilde G^{ST}\vert\ell
JM\rangle\Bigr\}\ .\end{eqnarray*}} \hfill\parbox{1cm}{
\begin{equation}\end{equation}}
Eq.~(26) was split into two
summations for the sake of faster computation as \hfill\break
\parbox{14.6cm}{\begin{eqnarray*} \tilde G^{LTS}_{k_Sq_S}({\bf k},
{\bf k}',{\bf P}\vert \vert {\bf k}+ {\bf k}')&=& \hat
L\sum_{J\ell\ell '}\imath^{\ell -\ell'}\hat
J^3(-1)^\ell\hat\ell\hat S\ \tilde G^{LTS}_{} (\ell 'J,\ell J,P)\\
& \times &\sum_{k_\ell q_\ell \Lambda }\hat k_\ell (-1)^{k_\ell}\
\left\{\matrix{\ell &\ell '&k_\ell\cr J&J&L\cr S&S&k_S
\cr}\right\}\\ & \times &
\langle k_\ell q_\ell ,L\Lambda \vert k_Sq_S\rangle\
\langle\ell ' q_\ell ,\ell 0\vert k_\ell q_\ell \rangle\\
& \times & Y_{\ell
'q_\ell}(\theta ,0)\ {Y_{L\Lambda}(\theta /2,0)} \
.\end{eqnarray*}}
\hfill\parbox{1cm}{\begin{eqnarray}\end{eqnarray}} When expressed
in terms of the components of the tensors $\tilde
G^{LTS}_{k_Sq_S}({\bf k},{\bf k}', {\bf P}\vert \vert {\bf k}+
{\bf k}')$ in the `standard' coordinate system, Eqs.~(19)-(22)
reduce to \begin{eqnarray}\tilde G^{LTS}_C(\theta )&=&\hat S^{-2}\
\tilde G^{LTS}_{00}\\ \tilde G^{L1T}_{LS}(\theta
)&=&{i\sqrt{3}\over 6}\ \tilde G^{L1T}_{11}\\ \tilde
G^{L1T}_{TD}(\theta )&=&{\sqrt{2}\over 72}\ \Bigl[ 3(1-\cos\theta
)\tilde G^{L1T}_{20}+2\sqrt{6} \sin\theta\ \tilde
G^{L1T}_{21}+\sqrt{6}(3+\cos\theta )\tilde G^{L1T}_{22}\Bigr]\\
 \tilde G^{L1T}_{TX}(\theta ) &=&{\sqrt{2}\over 72}\ \Bigl[
3(1+\cos\theta )\tilde G^{L1T}_{20}-2\sqrt{6}\sin\theta\ \tilde
G^{L1T}_{21} +\sqrt{6}(3-\cos\theta )\tilde G^{L1T}_{22}\Bigr]\
,\end{eqnarray} where the $\theta $ arguments  of the tensor
components on the right-hand side have been dropped for
simplicity.

An additional transform was needed at the end to replace states
where $S=0,1$ and $T=0,1$ with the singlet and triplet spin and
isospin operators customarily used by the distorted-wave programs.

In Ref.~\cite{r10}, it was noted that, if the $M$-dependent
$G$-matrix elements were simply averaged over $M$, the result was
very close to the one obtained with the standard angle-average
calculation. The amplitudes obtained here by retaining only the
$L=0$ term  were also observed to reproduce the spherically
averaged results to excellent precision.  This was used as a check
of the computational algorithms.

Values of the density-dependent $G$-matrix calculated with the
spherical Pauli operator were used for partial waves where $7\leq
J\leq 15$, matrix elements that we needed to specify the
long-range pion tail of the NN interaction.  No partial waves with
$J>15$ were included. The amplitudes of Eqs.~(29)--(32) were then
calculated at a number of values of $\theta$ and reproduced using
a sum of Yukawa functions [5,14]. Because the $L=2$ contributions
were much smaller than for $L=0$, only $L=2$ was considered and
terms with $L\geq 4$ were ignored.

The matrix inversion scheme described in Appendix A of
Ref.~\cite{r5} presumes that the amplitudes associated with the
$S_{12}(\hat q)$ and $S_{12}(\hat Q)$ are related according to
$\tilde G^{LST}_{TD}(\pi -\theta )= -(-)^{S+T}\ \tilde
G^{LST}_{TX}(\theta )$ (as reported for $E^\prime$ and $F^\prime$
in Love and Franey \cite{r14}). For our present situation in which
{\bf P} is chosen to be parallel to $(\bf{k}+{\bf k}')$, this
relationship is no longer valid (see the discussion in Appendix
B).  So we adopted an older fitting scheme in which the $\tilde
G^{LST}_i(\theta )$ are calculated as a function of scattering
angle and reproduced using standard least squares minimization
techniques to determine the Yukawa expansion coefficients.
Separate coefficients were obtained for $\tilde
G^{LST}_{TD}(\theta )$ and $\tilde G^{LST}_{TX}(\theta )$, leading
to the same quality of reproduction usually obtained for the
spherical Pauli blocking case.

\section{Results for (p,p$'$) reactions}

In the framework referred to by the name of Brueckner-Hartree-Fock
(BHF), density-dependent effects on the interaction are included
through Eq.~(2), with the energy denominator properly modified by
the presence of the nuclear matter mean field. Considerations of
nucleons moving through nuclear matter as Dirac particles result
in what is referred to as the Dirac-Brueckner-Hartree-Fock (DBHF)
approach \cite{r24}. The latter is known to provide a realistic
description of the nuclear matter equation of state \cite{r18}.

\begin{figure}[p]
\center
\includegraphics[height=16.5cm]{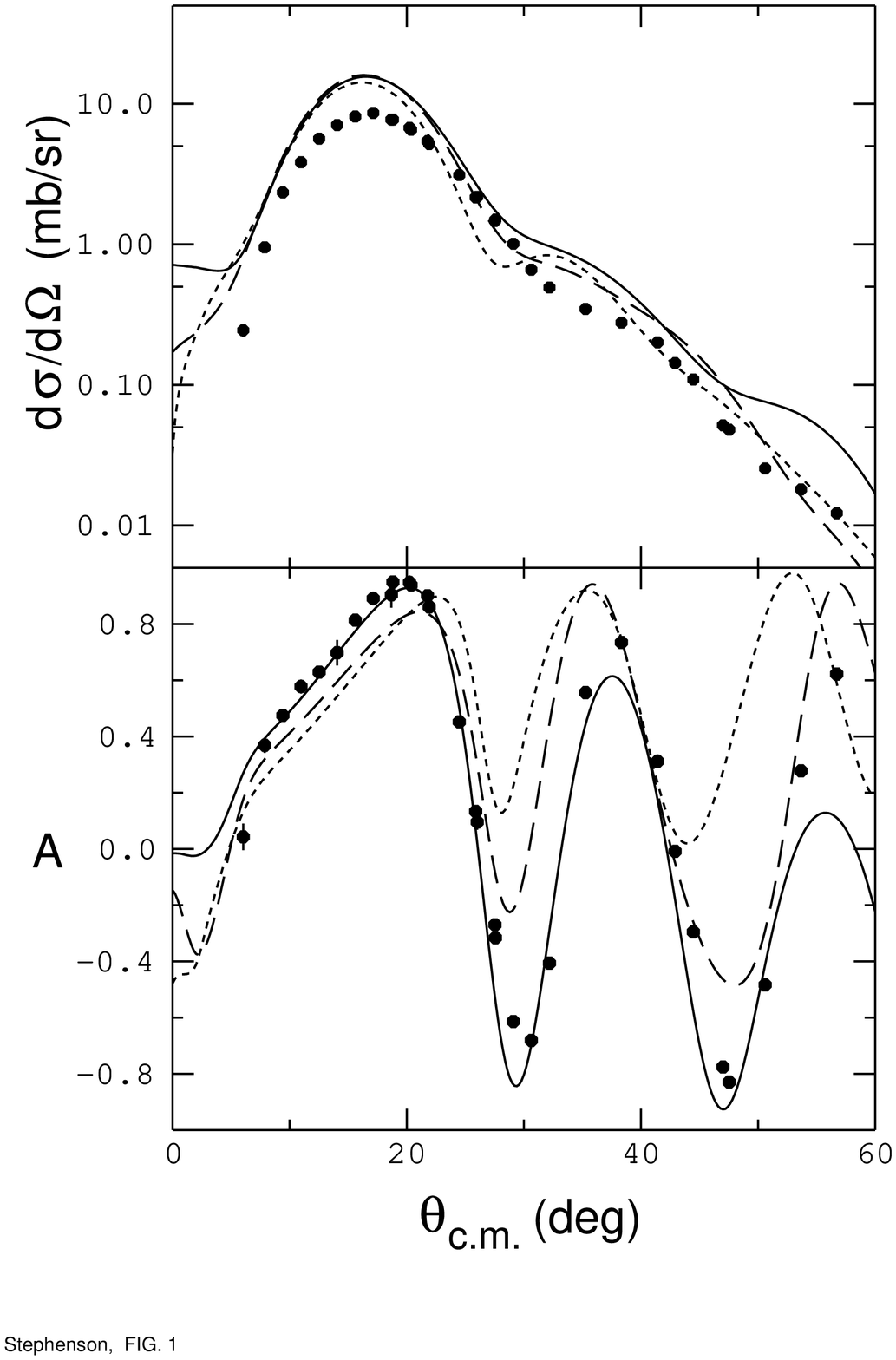}
\caption{Measurements at 200~MeV of the cross section and
analyzing power for the $^{40}$Ca(p,p$'$) transition to the $3^-$
state at 3.736~MeV.  The curves are DWIA calculations that include
no medium effects (short dash), spherical BHF effects (long dash),
and DBHF effects (solid).}
\end{figure}

It is useful to gauge the difference between the spherical and
non-spherical Pauli operators against the background of these
other contributions to medium modifications.  To illustrate their
effects, in Fig.~1 we show the cross section and analyzing power
data for the $3^-$ state at 3.736~MeV in $^{40}$Ca \cite{r25}. The
beam energy for these measurements was 200~MeV.  The short dashed
curves contain no density dependence.  Including only BHF effects
produces the long-dashed curves; DBHF calculations produce the
solid curves.  As discussed in Ref.~\cite{r5}, the distortions are
based on a folded optical potential that was constructed using the
same effective interaction as the DWIA transition.  The local
nuclear matter density was unfolded from the longitudinal electron
scattering form factor \cite{r26} using the proton charge
distribution. The structure formfactors replicate the inelastic
electron scattering measurements for this same transition
\cite{r25}. The calculations were made with the zero-range program
LEA \cite{r27}. With the treatment of exchange discussed in
Appendix~B of Ref.~\cite{r5}, the zero-range approximation
provides an adequate description of natural parity transitions, as
shown by the comparisons illustrated in Fig.~3 of Ref.~\cite{r5}.
Using this method, we can maintain a high quality treatment of the
transition formfactor.

The most striking effects of the increasing repulsion in the
nuclear medium lie beyond $20^\circ$ where they drive the
analyzing power to more negative values and increase the cross
section.  With DBHF, very good agreement to the analyzing power
data is obtained between $8^\circ$ and $35^\circ$ where the cross
section is the largest and the DWIA should be at its best.
However, the cross section is overestimated by a factor of about
2, a problem of unknown origin with the DWIA that was circumvented
for an {\it empirical} effective interaction by simply decreasing
the normalization of the NN amplitudes at all densities
\cite{r25}. If the calculated cross sections are lowered in this
manner to agree with the data at the cross section peak, the DBHF
curve provides the best reproduction of the shape of the angular
distribution.

\begin{figure}[p]
\center
\includegraphics[height=16.5cm]{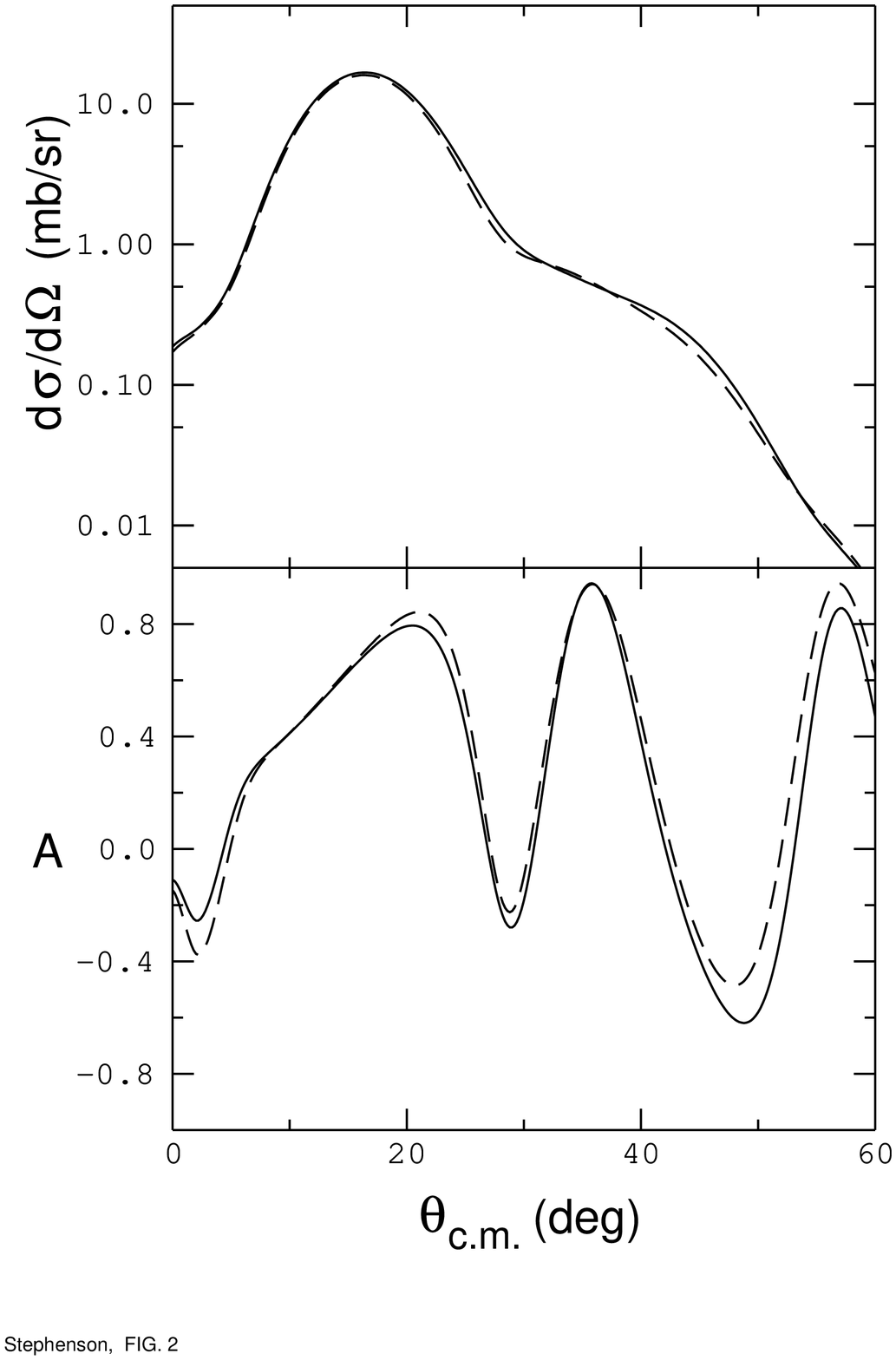}
\caption{Comparison of DWIA calculations with spherical (dashed)
and non-spherical (solid) Pauli density dependence.  The
calculations at 200~MeV are for the $^{40}$Ca(p,p$'$) transition
to the $3^-$ state at 3.736~MeV.}
\end{figure}

In Fig.~2 we show for the same transition BHF-based predictions
with non-spherical (solid) and spherical (dashed) Pauli operators.
(At this time we do not have DBHF calculations with non-spherical
Pauli blocking.) The differences are modest at all angles, being
no more than a small fraction of the changes shown in Fig.~1.  So
given the quality of agreement with the DBHF calculations from
Fig.~1, it is not possible to conclude whether the addition of
non-spherical components to the treatment of Pauli blocking is
required by the data.

\begin{figure}[p]
\center
\includegraphics[height=16.5cm]{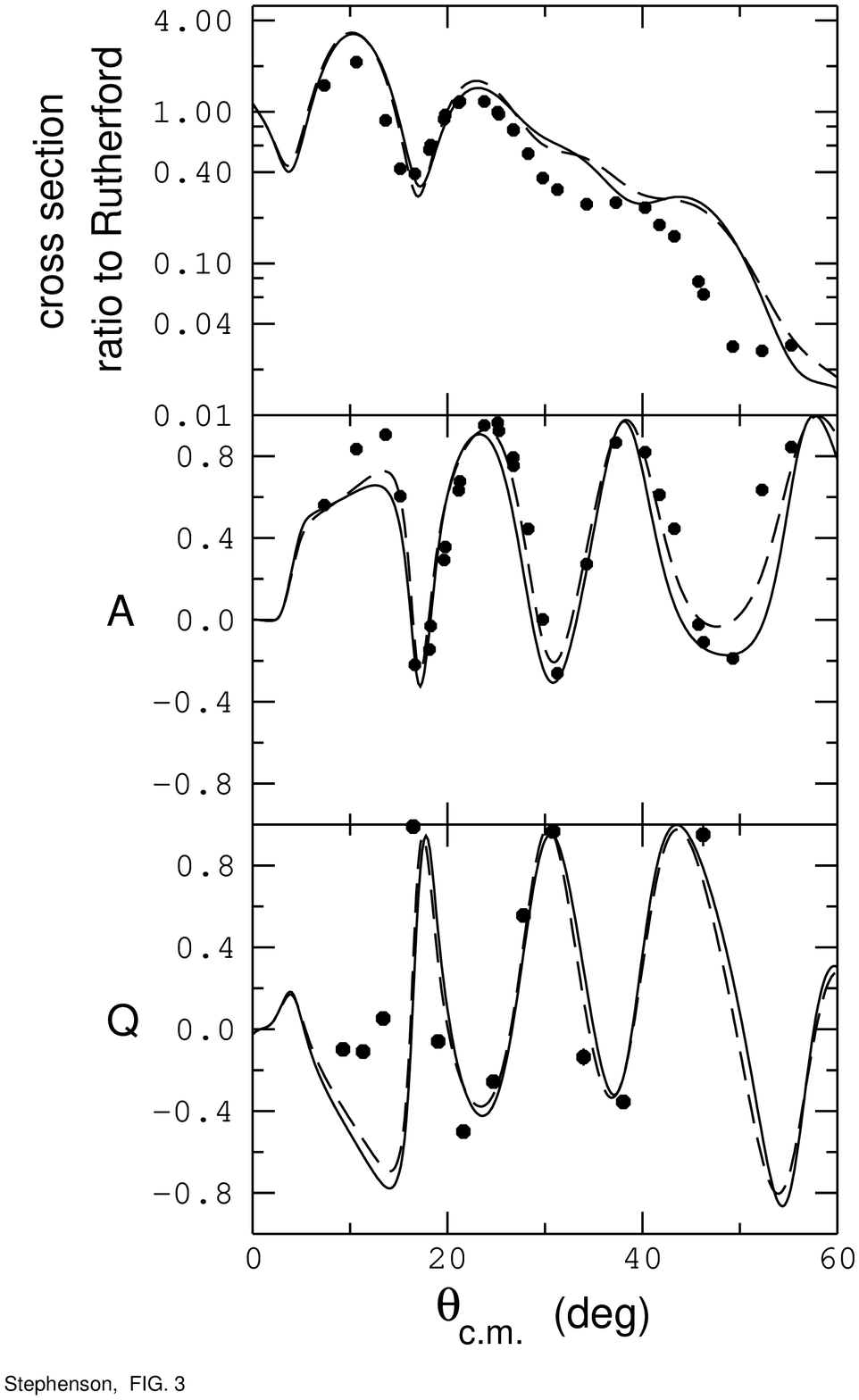}
\caption{Measurements of the cross section, analyzing power, and
spin rotation parameter $Q$ at 200~MeV for proton elastic
scattering from $^{40}$Ca.  The folded optical model calculations
are based on density-dependent interactions that include either
spherical (dashed) or non-spherical (solid) Pauli effects.}
\end{figure}

In Fig.~3, the non-spherical (solid) and spherical (dashed) BHF
calculations are compared with measurements of the cross section
\cite{r25}, analyzing power \cite{r25}, and spin rotation
parameter $Q(\theta )$ \cite{r28} for proton elastic scattering
from $^{40}$Ca at 200~MeV. To reduce complication, calculations
without density dependence and DBHF calculations are not shown.
Our earlier results \cite{r5} demonstrate that best agreement with
elastic scattering is often obtained with just the BHF density
dependence, a conclusion that has historically been verified many
times [4]. Finite nucleus effects substantially reduce the size of
the relativistic medium effects for elastic scattering \cite{r29}.
This reduction does not take place when the density dependence is
obtained through calculations in infinite nuclear matter.  Thus
agreement between these BHF calculations and elastic scattering is
good for both $A(\theta )$ and $Q(\theta )$ in the middle of the
angular range of the data. Relativistic effects at angles below
$15^\circ$, when included, do increase the analyzing power and
shift the spin rotation parameter toward more positive values.
Thus these differences where the elastic cross section is large
can be reduced in the fuller treatment of the medium.  The
differences between the Pauli operators with and without
non-spherical components are again too small to have an impact on
the quality of the agreement with experimental data.

\begin{figure}[p]
\center
\includegraphics[height=16.5cm]{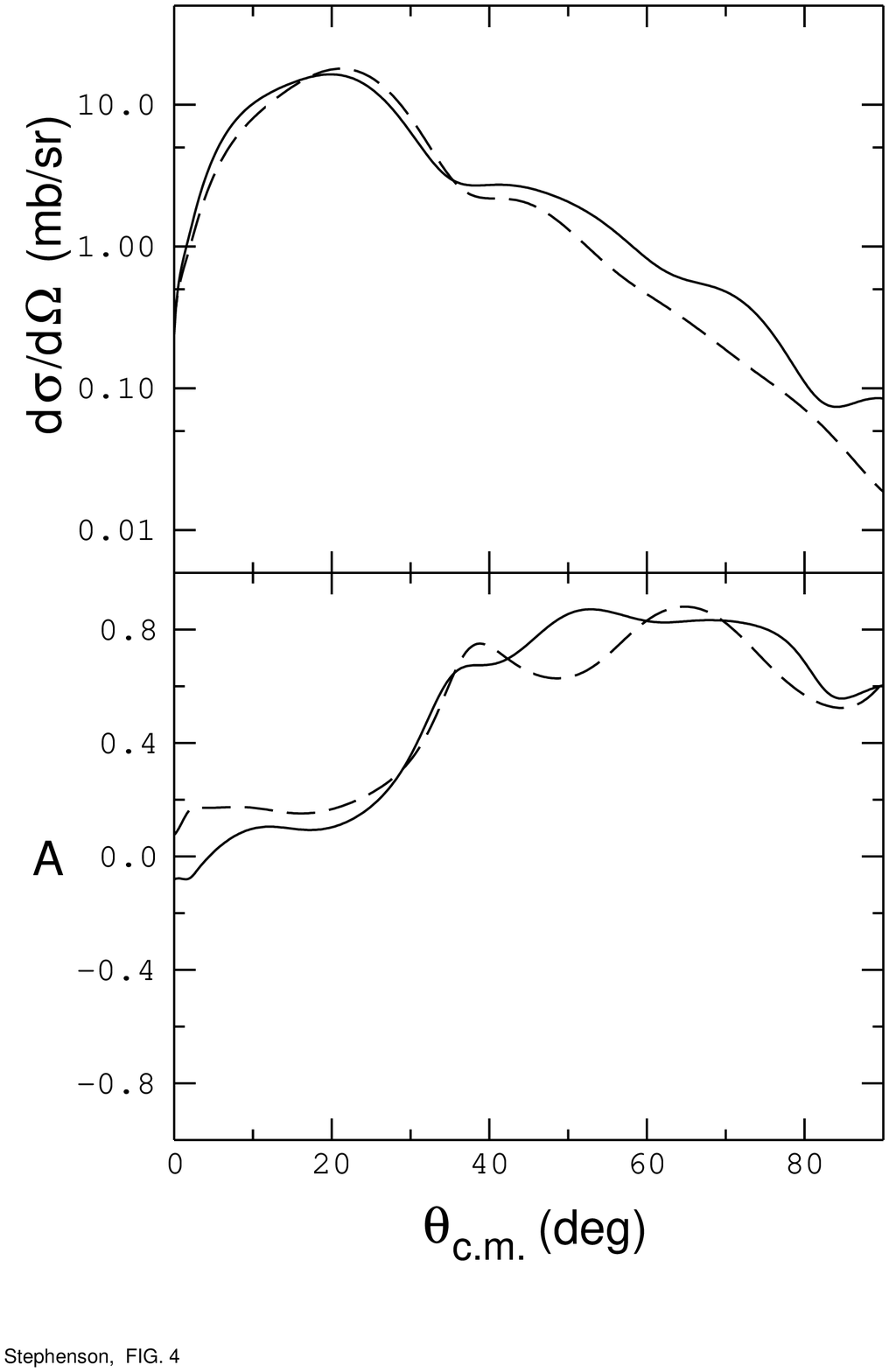}
\caption{Comparison of DWIA calculations with spherical (dashed)
and non-spherical (solid) Pauli density dependence.  The
calculations at 100~MeV are for the $^{40}$Ca(p,p$'$) transition
to the $3^-$ state at 3.736~MeV.}
\end{figure}

Last, it is known that Pauli blocking effects become weaker as the
bombarding energy goes up, finally disappearing near the pion
production threshold.  Figure~4 shows the non-spherical (solid)
and spherical (dashed) BHF calculations for the $3^-$ state in
$^{40}$Ca at 100~MeV.  As expected, the differences are larger
than in Fig.~2, but still remain smaller than the changes
associated with the baseline medium effects or a DBHF treatment.
(Earlier calculations \cite{r30} that showed much larger effects
at 100~MeV were in error.)

\section{Conclusions}

A number of authors have examined the non-spherical Pauli blocking
operator and calculated the changes to the $G$-matrix elements
compared to the spherical approximation.  We report here for the
first time a formalism that allows the effects of the
non-spherical operator to be included in a calculation of
nucleon-induced elastic and inelastic scattering.  We truncate the
resulting $G$-matrix to remove the elements with $J\neq J'$ or
$\vert\ell -\ell '\vert
>2$. This truncation involves terms that are small relative to the
changes in the $G$-matrix elements with $M$, the projection
quantum number for $J$.  An expansion is introduced in terms of
${\bf L}={\bf J}-{\bf J}'$ where the $L=0$ part essentially
recovers the spherically-averaged Pauli blocking result.  In this
expansion we restrict our consideration to terms with $L<4$.  We
wished to look in particular at polarization observables that
might be sensitive to the $M$-dependence of the interaction matrix
elements.

Brueckner-Hartree-Fock calculations of the effective interaction
including the non-spherical Pauli operator were made at 100 and
200~MeV. They were compared against other BHF-predictions obtained
with the more common angle-average approximation, as well as DBHF
calculations and those that contained no density dependence at
all. The change from spherical to non-spherical in the treatment
of the Pauli blocking operator produced changes that were small in
comparison to the effects associated with typical BHF or DBHF
medium modifications. Given the experimental errors and the likely
larger uncertainties in the DWIA theory, the small size of the
changes associated with adopting the non-spherical Pauli treatment
cannot be deemed significant. These conclusions remained valid
with changing bombarding energy. No polarization observables (many
were examined) were found to be particularly sensitive to the
non-spherical treatment of the Pauli operator.  Thus at the level
at which present theory can provide an accurate model of
nucleon-induced reactions, we find the spherically-averaged
treatment of the Pauli projection operator adequate for
intermediate-energy proton-nucleus scattering.

\vspace{0.5cm}

The authors acknowledge financial support from the U.S. National
Science Foundation under grant number PHY-0100348 (E.J.S.), the
U.S. Department of Energy under grant number DE-FG02-03ER41270
(F.S.) and the U.K. Engineering and Physical Sciences Research
Council under grant number GR/M82141.

%\section*{Appendix A}
\appendix
\section{Symmetries of the $G$-matrix in infinite uniform nuclear
matter}

We define the $G$-matrix as the solution of Eq.~(2) and  think of
it as the set of matrix elements of an operator in the NN relative
coordinate space for a fixed ${\bf P}$ and $k_F$. This operator
does not conserve the total angular momentum in the NN
center-of-mass if ${\bf P}\neq 0$. However, the Pauli blocking
operator $Q$ is invariant under spatial reflections and time
reversal and we will assume that the NN interaction $V$ is also.
These symmetries place important restrictions on the free-space
$G$-matrix. Here we examine their consequences  for the structure
of the on-shell $G$-matrix in nuclear matter.

  If the two-body interaction $V$ is invariant under time reversal, it is
straight forward to show formally from Eq.~(2) that the $G ({\bf
P})$  operator for fixed ${\bf P}$ satisfies \begin{equation}K
G({\bf P})K ^{-1}=G({\bf P})^{\dagger}\ ,\end{equation} where $K$
is the anti-unitary time reversal operator for the system. What
this relationship means is that for any states $\vert \phi\rangle$
and $\vert \psi\rangle$
 \begin{equation}\langle \phi\mid G({\bf P})\mid \psi\rangle
=\langle K \psi\mid G({\bf P})\mid  K\phi\rangle,\end{equation}
where $\vert K\psi\rangle\equiv K\mid \psi\rangle$.

  Our angular momentum states are chosen to have phases such that
\begin{equation}K \mid (lS)JM\rangle=(-1)^{J+M+\ell}\mid
(lS)J-M\rangle\ .\end{equation} The occurence of the phase
$(-1)^{\ell }$  in Eq.~(35) implies that we have not included
$\imath ^{\ell}$ factors with the orbital angular momentum
wavefunctions $Y_{\ell m}$. This is consistent with what is
assumed in Eq.~(10).

 Using the results of Eqs.~(34) and (35) we obtain\hfill\break
\parbox{14.6cm}{\begin{eqnarray*}\lefteqn{
\langle (l^{\prime}S)J^{\prime}M^{\prime}
 \mid G^{T}({\bf P})\mid (lS)JM \rangle}\\ & = &
(-1)^{\ell'+J^{\prime}+M^{\prime}+J+\ell +M}\langle (lS)J-M
 \mid G^{T}({\bf P})\mid (l^{\prime}S)J^{\prime}-M^{\prime} \rangle\ .
\end{eqnarray*}}\hfill\parbox{1cm}{\begin{equation}\end{equation}}

In the case of no Pauli blocking, $G$  conserves $J$ and $M$ and
the $G$-matrix is independent of $M$. Together with parity
conservation, this leads to the usual result that the $G$-matrix
is symmetric in the angular momentum basis.

When Eq.~(36) is used in the definition of $G^{LTS}$ from Eq.~(9),
we find
\begin{equation}
\tilde G^{LTS}(l^{\prime}J^{\prime},lJ,P)= (-1)^{J^{\prime}+\ell
'-J-\ell }{\hat{J}^{\prime}\over \hat{J}}\tilde
G^{LTS}(lJ,l^{\prime}J^{\prime},P)\ , \end{equation} and using
this in Eq.~(25) (assuming $k=k'$) gives
\begin{equation}
\tilde G^{LTS}_{k_{S}q_{S}}({\bf k},{\bf k}^{\prime},{\bf
P})=(-1)^{k_S}\tilde G^{LTS}_{k_{S}q_{S }}(-{\bf k}^{\prime},-{\bf
k},{\bf P})\ . \end{equation} Eq.~(38) is valid for arbitrary
fixed ${\bf P}$.

For fixed ${\bf P}$ a consequence of parity conservation in
Eq.~(25) is
\begin{equation}
\tilde G_{k_{S}q_{S}}^{LTS}({\bf k},{\bf k}^{\prime},{\bf
P})=\tilde G_{k_{S}q_{S}}^{LT S}(-{\bf k},-{\bf k}^{\prime},{\bf
P})\ ,\end{equation} Taken together, Eqs.~(38) and (39) imply
\begin{equation}
\tilde G_{k_{S}q_{S}}^{LTS}({\bf k},{\bf k}^{\prime},{\bf
P})=(-1)^{k_S}\tilde G_{k_{S}q_{S }}^{LT S}({\bf k}^{\prime},{\bf
k},{\bf P})\ . \end{equation}
 Eqs.~(38) and (39) are the generalization of  the usual
prescriptions for constructing $G$.  They lead to  the form given
in Eq.~(15) in the spherical Pauli-blocking  approximation. For a
general ${\bf P}$, Eqs.~(38) and (39) do not restrict $G$ to have
the form of Eq.~(15) and
 there will be many other terms involving ${\bf P}$,
for example $S_{12}({\bf P})$.

We can look at the consequences of Eqs.~(38) and (39) in a
different way. Eq.~(40) is a relationship between two different
sets of $G$-matrix elements corresponding to different ingoing and
outgoing momenta with the same (arbitrary) ${\bf P}$. We can
change this into a relation between matrix elements with the same
ingoing and outgoing momenta but  with different ${\bf P}$'s. We
do this by noting that, if $k=k'$ interchanging ${\bf k}$ and
${\bf k'}$ (as in Eq.~(40)) is the same as rotating them by $\pi$
about ${\bf k}+{\bf k'}$. Eq.~(40) is therefore equivalent to
\begin{equation}
\tilde G_{k_{S}q_{S}}^{LTS}({\bf k},{\bf k}^{\prime},{\bf
P})=(-1)^{k_S}\tilde G_{k_{S}q_{S }}^{LT S}(R{\bf k},R{\bf
k}^{\prime}, R{\bf P}^{\prime})\ , \end{equation} where $R$
denotes the rotation by $\pi$ about ${\bf k}+{\bf k'}$ and
\begin{equation}{\bf P}^{\prime}=R^{-1}{\bf P}\ .\end{equation}
We next use the fact that the $\tilde G_{k_Sq_S}^{TS}$ are
components of an irreducible tensor of rank $k_S,q_S$. Therefore
the condition  Eq.~(41), can be written \cite{r31}
\begin{equation}
\tilde G_{k_{S}q_{S}}^{LTS}({\bf k},{\bf k}^{\prime},{\bf P})
)=(-1)^{k_S}\sum_{q^{\prime}_{S}}\tilde
G_{k^{}_{S}q^{\prime}_{S}}^{LT S}({\bf k},{\bf k}^{\prime},{\bf
P}^{\prime}) {\mathcal D}^{k_{S}}_{q^{\prime}_{S},q^{}_{S}}({\bf
k}+{\bf k'},\pi )\ , \end{equation} where ${\mathcal
D}^{k_{S}}_{q^{\prime}_{S},q^{}_{S}}({\bf k}+{\bf k'},\pi )$ is
the rotation matrix corresponding to a rotation of $\pi $ about
${\bf k}+{\bf k'}$.

A similar argument based on the parity condition, Eq.~(39), and
the fact that $-{\bf k}=R(\pi ,{\bf n}){\bf k}$ and $-{\bf
k}^{\prime}= R(\pi ,{\bf n}){\bf k}^{\prime}$ gives \cite{r31},
\begin{equation}
\tilde G_{k_{S}q_{S}}^{LTS}({\bf k},{\bf k}^{\prime},{\bf P})
)=(-1)^{k_S}\sum_{q^{\prime}_{S}}\tilde
G_{k^{}_{S}q^{\prime}_{S}}^{LT S}({\bf k},{\bf k}^{\prime},{\bf
P}^{\prime \prime}) {\mathcal
D}^{k_{S}}_{q^{\prime}_{S},q^{}_{S}}({\bf n},\pi )\ ,
\end{equation} where ${\bf P}^{\prime \prime}=R^{-1}(\pi, {\bf
n}){\bf P}$.

In free space, or in the spherical Pauli blocking approximation,
the $G$-matrix does not depend on the direction of ${\bf P}$ and
Eqs.~(43) and (44) become linear relationships between tensor
components $\tilde G_{k_{S}q_{S}}^{LTS}$ of the same rank.
Expressed in terms of components in the ``standard'' coordinate
system, Eq.~(44) (parity) gives
\begin{equation}
\tilde G_{k_{S}q_{S}}^{LTS}=(-1)^{k_S+q_S}\tilde G_{k_{S}-q
_{S}}^{LT S}. \end{equation}
 This simply states that for $k_S=1$ the vector 
$\tilde G_{1q_{S}}^{LTS}$ is
proportional to ${\bf n}$ and that the $k_S=2$ tensor is
determined by 3 complex amplitudes instead of the general 5:
\hfill\break \parbox{14.6cm}{\begin{eqnarray*} & \tilde
G^{LST}_{10}=0,\ \ \ \tilde G^{LST}_{11}=\tilde G^{LST}_{1-1} & \\
&\tilde G^{LST}_{20},\ \ \ \tilde G^{LST}_{21}=-\tilde
G^{LST}_{2-1},\ \ \tilde G^{LST}_{22}=\tilde G^{LST}_{2-2}\ . &
\end{eqnarray*}}\hfill\parbox{1cm}{\begin{eqnarray}\end{eqnarray}}
  Hence  for $S=1$ a consequence of Eq.~(45) in free space is that the
$G$-matrix is determined by 5 complex $\theta$-dependent
amplitudes.

Equation (43) (or Eq.~(38), time reversal) gives no extra
information for $k_S=1$, but for $k_S=2$ gives the extra condition
\begin{equation}
\sin \theta (\sqrt{3/2}\ \tilde G^{LST}_{20}-\tilde
G^{LST}_{22})+2\cos \theta \,\tilde G^{LST}_{21}=0\
,\end{equation}
 so that in fact there are only four independent $G$-matrix amplitudes. (That
Eq.~(47) is the only extra condition can easily be seen by
expressing Eq.~(44) in a coordinate system with z along ${\bf
k}+{\bf k'}$. It is found that Eq.~(44) implies $\tilde
G^{LST}_{21}=0$ in this system \cite{r31}.)   This is the content
of the form given in Eq.~(15).

  For a general ${\bf P}$ in nuclear matter none of these restrictions will be
true because Eqs.~(43) and (44) relate tensor components with
different ${\bf P}$. However, the $G$-operator satisfies $\tilde
G(-{\bf P})=\tilde G({\bf P})$ and hence if
  \begin{equation}{\bf P}^{\prime \prime}=\pm {\bf P}
\end{equation}
and
 \begin{equation}{\bf P}^{\prime}=\pm {\bf P}\ ,\end{equation}
where ${\bf P}^{\prime}$ and ${\bf P}^{\prime \prime}$ are defined
in Eq.~(42) and following Eq.~(44), respectively,  then
Eq.~(44) does relate tensor components with the same {\bf P} and
Eqs.~(46) and (47), will be satisfied.

This situation occurs when ${\bf P}$ lies in the scattering plane;
both Eqs.~(48) and (49) are satisfied if ${\bf P}$ is parallel
to ${\bf k}+{\bf k'}$. This is the choice we make in this paper.
With this choice the conditions, Eqs.~(46) and (47), are
satisfied and, equivalently,
 the $G$-matrix has the form of Eq.~(15) on shell.

%\section*{Appendix B}

\section{Direct and exchange tensor amplitudes for arbitrary P}

We can ensure that the amplitudes $\tilde G^{LST}_i(\theta )$ of
Eq.~(15) are anti-symmetrized by subtracting in each case the same
amplitude with the additional phase factor $(-)^{S+T}$ and with
{\bf k} replaced by $-${\bf k}. This requires that any value of
$\ell$ that appears in a partial wave decomposition of $\tilde
G^{LST}_i(\theta )$ satisfies $(-)^{\ell +S+T}=-1$.

The direction of {\bf P}, the total momentum of the colliding
nucleons defined in Eq.~(1), remains the same for both terms in
the anti-symmetrization subtraction.  If we consider just the
amplitudes associated with the tensor spin operators $S_{12}(\hat
q)$ and $S_{12}(\hat Q)$, then in terms of the unsymmetrized
amplitudes $G^{LST}_i(\theta )$: \begin{equation} \langle{\bf
k'}\vert G({\bf P})\vert{\bf k}\rangle =  G^{LST}_{TD}(\cos\theta
, {\bf P}\kern-2pt\cdot\kern-2pt{\bf q},{\bf
P}\kern-2pt\cdot\kern-2pt{\bf Q}, {\bf
P}\kern-2pt\cdot\kern-2pt{\bf n})\ S_{12}(\hat q)+
G^{LST}_{TX}(\cos\theta , {\bf P}\kern-2pt\cdot\kern-2pt{\bf
q},{\bf P}\kern-2pt\cdot\kern-2pt{\bf Q}, {\bf
P}\kern-2pt\cdot\kern-2pt{\bf n})\ S_{12}(\hat Q)\end{equation}
where the dot products indicate any scalar generated with the
vector {\bf P}. The amplitude to be subtracted is\hfill\break
\parbox{14.6cm}{\begin{eqnarray*}\lefteqn{
\langle{\bf k'}\vert G({\bf P})\vert -{\bf k}\rangle =
G^{LST}_{TD}(-\cos\theta ,{\bf P}\kern-2pt\cdot\kern-2pt{\bf Q},
{\bf P}\kern-2pt\cdot\kern-2pt{\bf q},-{\bf
P}\kern-2pt\cdot\kern-2pt{\bf n}) \ S_{12}(\hat Q)}\\ & + &
G^{LST}_{TX}(-\cos\theta ,{\bf P}\kern-2pt\cdot\kern-2pt{\bf Q},
{\bf P}\kern-2pt\cdot\kern-2pt{\bf q},-{\bf
P}\kern-2pt\cdot\kern-2pt{\bf n}) \ S_{12}(\hat q)\
.\end{eqnarray*}} \hfill\parbox{1cm}{\begin{equation}\end{equation}}
Note that the definition of the scattering angle
becomes the complement of the original value. With the reversal of
the direction of {\bf k}, the vectors {\bf q} and {\bf Q} have
interchanged places and magnitudes while {\bf P} remains fixed.
Thus, the direction of $\hat n=\hat q\times\hat Q$ reverses.

By combining terms, the anti-symmetrized form $G^{LST}_i(\theta )$
is given by \begin{equation}\tilde G^{LST}_{TD}(\theta )=
G^{LST}_{TD}(\cos\theta , {\bf P}\kern-2pt\cdot\kern-2pt{\bf
q},{\bf P}\kern-2pt\cdot\kern-2pt{\bf Q}, {\bf
P}\kern-2pt\cdot\kern-2pt{\bf n})-(-)^{S+T}\
G^{LST}_{TX}(-\cos\theta , {\bf P}\kern-2pt\cdot\kern-2pt{\bf
Q},{\bf P}\kern-2pt\cdot\kern-2pt{\bf q}, -{\bf
P}\kern-2pt\cdot\kern-2pt{\bf n})\end{equation} for the
coefficient of $S_{12}(\hat q)$ and \hfill\break
\parbox{14.6cm}{\begin{eqnarray*}\tilde G^{LST}_{TX}(\theta )&=&
G^{LST}_{TX}(\cos\theta , {\bf P}\kern-2pt\cdot\kern-2pt{\bf
q},{\bf P}\kern-2pt\cdot\kern-2pt{\bf Q}, {\bf
P}\kern-2pt\cdot\kern-2pt{\bf n})\\ & & -(-)^{S+T}\
G^{LST}_{TD}(-\cos\theta , {\bf P}\kern-2pt\cdot\kern-2pt{\bf
Q},{\bf P}\kern-2pt\cdot\kern-2pt{\bf q}, -{\bf
P}\kern-2pt\cdot\kern-2pt{\bf n})\\ & & -(-)^{S+T}\ \Bigl[
G^{LST}_{TD}(-\cos\theta ,{\bf P}\kern-2pt\cdot\kern-2pt{\bf Q},
{\bf P}\kern-2pt\cdot\kern-2pt{\bf q},-{\bf
P}\kern-2pt\cdot\kern-2pt{\bf n})\\ & & -(-)^{S+T}\
G^{LST}_{TX}(\cos\theta ,{\bf P}\kern-2pt\cdot\kern-2pt{\bf q},
{\bf P}\kern-2pt\cdot\kern-2pt{\bf Q},{\bf
P}\kern-2pt\cdot\kern-2pt{\bf n}) \Bigr]\end{eqnarray*}}\hfill
\parbox{1cm}{\begin{eqnarray}\end{eqnarray}}
for the coefficient of $S_{12}(\hat Q)$.

For NN scattering, the anti-symmetrized amplitudes satisfy
\cite{r14} \begin{equation}\tilde G^{LST}_{TD}(\pi -\theta
)=-(-)^{S+T}\ \tilde G^{LST}_{TX}(\theta )\ . \end{equation} This
permits the same Yukawa expansion coefficients to be used for
$\tilde G^{LST}_{TD} (\theta )$ and $\tilde G^{LST}_{TX}(\theta )$
provided that $Q$ is used in place of $q$ in the Yukawa expansion
of the $TX$ term \cite{r14}. For the non-spherical calculations
used in this paper, {\bf P} is along the direction of $\hat Q$,
thus the dot products involving {\bf Q} in Eqs.~(52) and (53)
remain. Due to the presence of different arguments in the $TD$ and
$TX$ amplitudes, Eq.~(54) is not necessarily satisfied. This is
the case
 for the calculations reported here.

% now the references. delete or change fake bibitem. delete next three
%   lines and directly read in your .bbl file if you use bibtex.

% \begin{figure}
% \caption{}
% \label{}
% \end{figure}

\end{document}